\documentclass[fleqn]{aa}  
\usepackage{supertabular,booktabs}  
\usepackage{graphicx}
\usepackage{xcolor}
\usepackage{hyperref}
\usepackage{txfonts}
\usepackage{calligra}
\usepackage{calrsfs}
\usepackage[T1]{fontenc}
\usepackage[mathcal]{eucal}
\usepackage{longtable}
\usepackage{multicol}

\usepackage[switch, modulo]{lineno}

\sfcode`A=1000 \sfcode`B=1000 \sfcode`C=1000 \sfcode`D=1000
\sfcode`E=1000 \sfcode`F=1000 \sfcode`G=1000 \sfcode`H=1000
\sfcode`I=1000 \sfcode`J=1000 \sfcode`K=1000 \sfcode`L=1000
\sfcode`M=1000 \sfcode`N=1000 \sfcode`O=1000 \sfcode`P=1000
\sfcode`Q=1000 \sfcode`R=1000 \sfcode`S=1000 \sfcode`T=1000
\sfcode`U=1000 \sfcode`V=1000 \sfcode`W=1000 \sfcode`X=1000
\sfcode`Y=1000 \sfcode`Z=1000

\newcommand{\Msun}{\hbox{M$_\sun$}}

\newcommand{\ssi}{\hbox{$\sigma_{\rm SI}\,$}}

\begin{document}

   \title{The large cores of dark matter and globular clusters in AS1063. 
   Possible evidence of self-interacting dark matter. Or not.}
   \titlerunning{GC \& DM cores in AS1063}
   \author{J.M. Diego \inst{1}\fnmsep\thanks{jdiego@ifca.unican.es}
    }      
   \institute{Instituto de F\'isica de Cantabria (CSIC-UC). Avda. Los Castros s/n. 39005 Santander, Spain}
 \abstract{
 Deep JWST images of AS1063 reveals tens of thousands of globular clusters in the galaxy cluster AS1063. When compared with the lensing model based on the same JWST data, the distribution of globular clusters traces closely the distribution of lensing mass (mostly composed of dark matter). Interestingly, both the distributions of dark matter and globular clusters have large  central cores. However the size of the core in the distribution of globular clusters is about half the size the core of the dark matter distribution. We argue that the standard cold dark matter and fuzzy dark matter models struggle to explain these large cores. Meanwhile, the self interacting dark matter with a velocity dependent cross section, combined with core stalling, offers a natural explanation to the existence of these cores if  $\sigma_{\rm SI}\approx 0.3$ cm$^2$ g$^{-1}$ for galaxy cluster halos. But we also discuss how the lack of hydrodynamical N-body simulations capable of resolving globular clusters in galaxy cluster scale halos, hinders the possibility of ruling out the standard non-collisional dark matter scenario. Future high-resolution hydrodynamical simulations of galaxy clusters, with several trillion particles, and containing over a hundred thousand globular clusters, can provide the insight needed to transform the epistemic nature of dark matter into an ontological one.  
   }

   \maketitle
%
\section{Introduction}

Among the biggest unsolved mysteries in modern cosmology, the nature of dark matter (DM) stands as one of the oldest and most intriguing ones. While the standard (i.e., non-collisional) cold DM (CDM) model  has been very successful at explaining observations from the largest scales probed by the Cosmic Microwave Background \citep{Bahcall1999,Planck2020a} to the much smaller scales involved in halo formation \citep{Tegmark2004}, a possible tension exists with observations at the smallest scales where CDM predicts a cusp at the center of the halos, but observations of dwarf galaxies and clusters reveals the presence of cores \citep{Tyson1998}. 
Feedback from supernova (SN) or central active galactic nuclei (AGN)
 is often referred to as a possible mechanism to transform the CDM predicted cusp into a core. But in dwarf galaxies dominated by DM, this mechanism may be insufficient to explain the relatively large cores found in these galaxies \citep{Moore1994}, and tidal forces may be needed to explain their morphology \citep{Fattahi2018}.  On the opposite extreme, in very massive halos, the cores of some observed galaxy clusters  are much larger than expectations from simulations, representing a similar challenge for CDM. Alternative models of DM naturally predict the presence of cores in all halos, from dwarf galaxies to galaxy clusters, hence offering an interesting alternative to CDM. In models of DM where the DM particle is ultralight ($m_{\rm DM}<10^{-21}$ eV), DM behaves as a quantum fluid with fluctuations in the density field at scales corresponding to the de Broglie wavelength of  a particle with mass $m_{\rm DM}$. This type of model is inspired by the axion model in quantum chromodynamics, which was proposed to solve the strong charge conjugation -- parity violation, and notably manifested by the incredibly small electric dipole moment of the neutron. In these models, for sufficiently small values of $m_{\rm DM}$, the associated de Broglie scale becomes astrophysical. These models predict a soliton core at the center of the halos that could explain the observed cores. Conflictive results in the literature sometimes favor this type of model  \citep{Amruth2023,Broadhurst2025,Hou2026}, and sometimes disfavor it  \citep{Irsic2017,Dalal2022,Benito2025}. Perhaps, one of the main challenges for this type of model is that in order to explain the observed properties of small halos (dwarf galaxies) and large halos (galaxy clusters), the DM particle can not have a unique mass, but a spectrum of DM particle masses may be needed, with no clear mechanism explaining how different halos get populated by different DM particles with different masses. Or perhaps new physics govern the dynamics and evolution of these particles, but this is highly speculative and with no real observational nor theoretical evidence to back it up. This type of model receives many names in the literature, but here we simply refer to them as $\psi$DM. An alternative model, which is also physically motivated, is self-interacting DM (or SIDM)  \citep{Spergel2000,Rocha2013}. Although both $\psi$DM and SIDM can also be considered as particular cases of CDM (since DM is non-relativistic), the DM particles in SIDM have a small, but non-negligible (and much larger than in the standard CDM model), probability of interacting with each other, characterized by the cross-section \ssi. For values of the cross section $\ssi\approx 1$ cm$^2$ g$^{-1}$, the probability that two DM particles interact in the center of galaxy clusters during a Hubble time is approximately 1. This is sufficient to remove the predicted cusp in CDM models and transform it into a large core. On the much smaller dwarf galaxies, this value of \ssi is too small to produce noticeable effects, so \ssi should be at least an order of magnitude larger. A model where \ssi depends on the mass of the halo (or more precisely, on the typical velocities of the interacting DM particles, with larger velocities in more massive halos) is quite natural and appears in other areas of physics. Moreover, a velocity dependent \ssi naturally solves a different observational characteristic of small vs large halos. 
Due to the increased probability of interaction in smaller halos, these can experience gravothermal collapse, or core collapse,  transforming the SIDM core into a cusp. More massive halos, such as galaxy clusters, do not experience gravothermal collapse during a Hubble time and hence can retain their cores to present time. In dwarf galaxies (often found orbiting around a larger halo), where cores have been observed and where gravothermal collapse should be taking place, the combined effect of feedback and tidal forces from the main halo may be sufficient to disrupt the cusps at the center of the relatively shallow gravitational potentials. \\

Additional observational tests that can reveal departures from CDM and be better explained with alternative DM models are of paramount importance. An area that has not attracted much attention in relation to DM is the study of globular clusters (GCs) in galaxy clusters, especially from the theoretical or simulation point of view. Given the very compact nature of GCs (a few parsec in size), close encounters between GCs are expected to be incredibly rare in galaxy clusters. Together with their modest masses ($10^5$--$10^7\, \Msun$), GCs are expected to behave as collisionless particles and respond only to the gravitational forces in the cluster, in the same way as DM particles do. Hence, the study of the distribution of GCs and DM in the same cluster may provide valuable clues about the nature of DM.  

N-body simulations have not studied the role that the large number of GCs in a galaxy cluster ($\sim 10^5$) plays on the distribution of dark matter near the center of the cluster halo. This is mainly due to technical limitations. Galaxy clusters, while often seen as the largest linear structures in the universe that reach hydrostatic equilibrium (i.e., virialized), are sufficiently complex that analytical solutions are not accurate enough, and one needs to rely on simulations. Especially, the more expensive hydrodynamical simulations that take into account the interplay between baryonic physics and the dominant DM.  
N-body simulations that can simultaneously simulate a massive $10^{15}\, \Msun$ halo \citep[the total mass of AS1063 is estimated to be almost $3\times10^{15}\, \Msun$][]{Williamson2011}, with a sufficiently large number of particles to resolve $10^5\, \Msun$ GCs are incredibly complex. Such a simulation would require  a particle mass of less than $\sim 10^4\, \Msun$, in order to resolve the smallest GCs with at least 10 particles, implying a total number of particles of at least $10^{12}$ (or $10^{16}\, \Msun$ total mass in the simulation box), equivalent to 8 Tb of memory just to store one snapshot. Simulations with more than a trillion particles have already been produced \citep{Maksimova2021}, so a simulation with the required resolution is within the realm of possibility. Only recently have N-body codes managed to simulate a $5\times10^{14}\, \Msun$ galaxy cluster with a stellar particle mass of $2\times10^4\, \Msun$, and a spatial resolution of 68 pc \citep{Han2026}, much better than the $\sim 1$ kpc resolution of other state-of-the-art simulations. But even in this case, the resolution may not be sufficient to properly resolve GCs. When compared with lower resolution simulations, no appreciable differences are observed in quantities such as the density profile \citep{Jeon2025}. 
High resolution simulations capable of resolving GCs exist on the much smaller scale of $10^{12}\, \Msun$ halos, but the number of GCs and mass of the halo is proportionally much smaller \citep{Pfeffer2018,ReinaCampos2023}. Also, in galaxy cluster scales, we expect some of the objects classified as GCs (basically unresolved sources in the galaxy cluster) to be the massive remnants of galactic cores that can survive strong tidal interactions in the cluster thanks to the super massive black hole (SMBH) at their center. This adds another complication to the simulation and extends the dynamical range of masses of the GCs that should include also the more massive galactic core remnants. To the best of the knowledge of the author, such simulation has not been carried out so far on galaxy cluster scales so we are still lacking the intuition and support provided by numerical simulations. Important effects such as dynamical friction or gravothermal collapse of the system of GCs have not been studied on galaxy cluster scales. The later is expected to be a smaller effect than in stellar clusters since the probability of two-body relaxation between GCs is much smaller (owing to their relatively low volumetric density, and hence very low probability of close encounters). As for dynamical friction, this could play an important role, although as noted by \cite{vandenBosch2025}, dynamical friction ceases to operate in cores of (roughly) constant density, a phenomenon known as core stalling \citep{Read2006,Goerdt2010,Petts2015,Kaur2022}.
Semianalytic approaches have been explored \citep{Park2022}, but still lacking the detail provided by true N-body hydrodynamical simulations.

Keeping this context in mind (a relatively unexplored territory), here we study the distribution of DM and the system of globular clusters (GC) found around the central region of the massive galaxy cluster AS1063, and discuss it in the context of CDM and alternative DM models. In \cite{Diego2026c} (or paper-I) we studied this cluster and derived a new lens model based on recent JWST data. AS1063 is among the best studied gravitational lenses with tens of lensed galaxies serving as lensing constraints. The distribution of mass in the lens model is dominated by the DM component, so the lens model correlates strongly with the DM distribution. AS1063 is also among the deepest observations carried out with JWST, thus allowing for a detailed view of the system of GCs around a massive galaxy cluster. In a second paper \citep[][or paper-II]{Diego2026d} we derive the distribution of GCs in AS1063, finding over 30000 GCs in the inner region, and study the correlation with the lens model. This large number of GCs represents only a fraction of the total number of GCs expected in AS1063, which is estimated to host over $10^5$ GCs \citep{Diego2026d}. In general we find very good agreement between the lens model and GC distribution, especially after convolving the distribution of GCs with a kernel that transforms the discrete GC distribution into a continuous one that resembles well the lensing convergence. In this work we pay more attention to the differences between the GC and DM distributions in the core region, and discuss them in the context of DM models.

Throughout the paper we adopt a standard flat cosmological model with $\Omega_M=0.3$ and $h=0.7$. At the redshift of the lens ($z=0.348$), and for this cosmological model in particular, one arcsecond corresponds to 4.921~kpc. 

\begin{figure*} 
  \includegraphics[width=18cm]{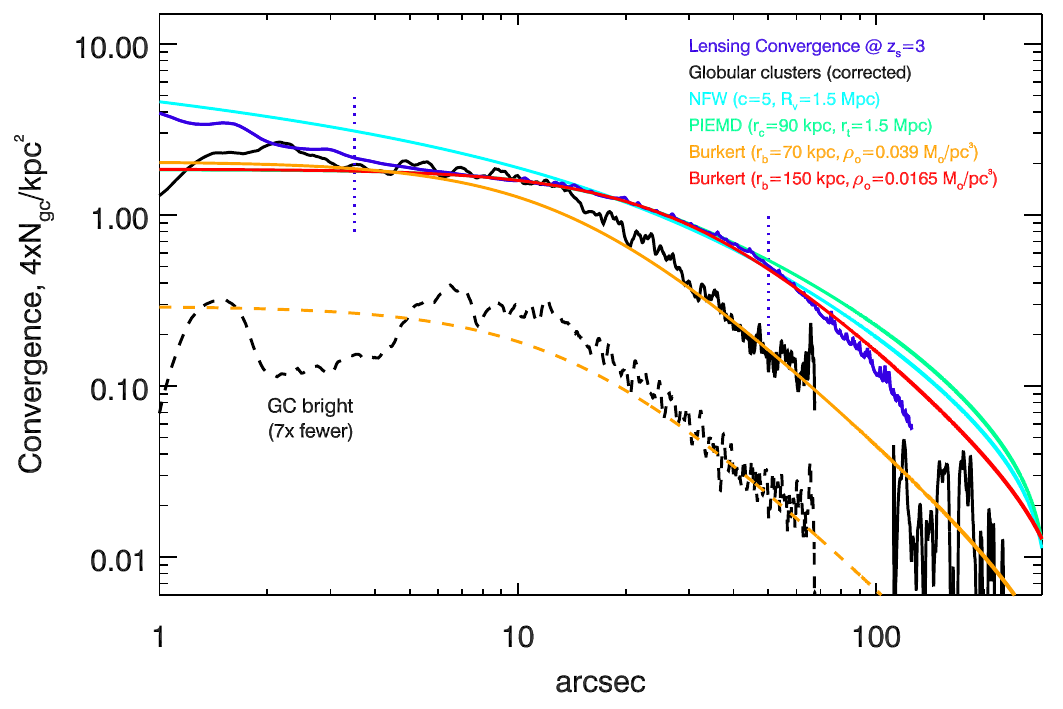} 
      \caption{Comparison of the lensing mass and GC profiles with analytical models. The dark blue (lensing mass) and black  (GCs) lines are from paper-I and paper-II respectively. For the GC, the profile is corrected for background contamination. The solid black curve is re-scaled by a factor 4 to visually match the lensing profile. The green curve is an NFW profile that matches the lensing profile in the region constrained by lensing (blue vertical dotted lines). The green line is a PIEMD profile fitting the lensing mass from alternative lens models (see text). The red line is a  Burkert profile fitting the lensing mass profile while the orange line is a different Burkert profile fitting the distribution of GCs. The dashed-black is the number density of the brightest GCs in the full sample obtained after increasing the detection threshold by one order of magnitude (and containing 4061 GCs, or one seventh the total number in the full catalog).  The dashed black line includes the same factor 4 as the black solid line. Finally, the dashed orange line is the same Burkert profile shown as a solid-orange line with $r_b=70$ kpc but divided by a factor seven. 
         }
         \label{Fig_Profiles_V4}
\end{figure*}

\section{The cores of the GC system and the DM distribution in AS1063}
\label{sec_cores}
In paper-II we discuss the similarities between the two-dimensional distributions and one-dimensional  profiles of GCs and the lensing model. This similarity can be exploited to infer the distribution of the invisible DM from the observed distribution of GCs, for instance, by convolving the discrete distribution of GCs with a kernel (see Eq.~2 in paper-II) to derive a two-dimensional distribution for the lensing mass. While the connection between GC and DM is an interesting application, it is also compelling to consider the differences between the GC and lensing mass distributions. We reproduce the lensing mass (or convergence) from paper-I (solid dark blue) and GC distribution from paper-II (solid black) in  Fig.~\ref{Fig_Profiles_V4}. The profile of the number density of GCs is derived from the main JWST module, centered on the cluster and covering radii $r<70"$. The second module covers distances $120" \lesssim r \lesssim 250"$. No data is available between $70"<r<120"$, resulting in a gap in the GC profile.  The profiles for the GC number density have been corrected by an estimation of the contamination from foreground galaxies as described in paper-II. At large radii we observe that both the lensing mass and GC number density profiles fall close to the expected $r^{-2}$ law in Navarro-Frenk-White \citep[][]{NFW} (NFW) profiles. Both profiles also show a relatively large core but the different core sizes is the most obvious difference between the profiles of the GC distribution and the lensing convergence. Different definitions of core sizes can be found in the literature, each one referring to a different profile definition. Here we use the Burkert profile \citep{Burkert1995} as our ruler to measure core sizes. In this model, the 3D density profile, $\rho(r)$, and core size $r_b$ are defined by,
\begin{equation}
\rho(r) = \frac{\rho_o}{(1+r/r_b)(1+(r/r_b)^2)}\, .
\end{equation}
Burkert profiles are appropriate for cored profiles and can be used in different scenarios. In our case we use this profile in the context of self-interacting dark matter (SIDM) models, where a core, especially in galaxy clusters that have not had enough time to experience core collapse, is usually expected. On the other hand, cuspy profiles are often described by the classic NFW profile. This type of profile is common in the context of the standard non-collisional CDM scenario. We use both, the Burkert and NFW profiles to fit for the lensing mass. The results is shown in Fig.~\ref{Fig_Profiles_V4} as a light blue curve for the NFW case, and the Burkert profile is shown as a red curve. If we ignore the central peak due to the stellar component of the BCG below $r=2"$, in the region constrained by lensing (marked with two vertical dotted lines), the Burkert profile fits the lensing convergence better than the NFW. With the Burkert profile  we measure a core radius of 150 kpc for the DM halo. As discussed in paper-I, the lens model profile (dark blue line) is biased low beyond $r\approx 50"$ due to lack of lensing constraints at these distances. We see how both the NFW and Burkert profiles are above the lens model at $r<50"$ which is consistent with this bias in the lens model at these radii. For comparison we add a third profile in green, a pseudo-isothermal mass distribution \citep[or PIEMD,][]{Kassiola1993,Kneib1996,Limousin2005} that correspond to two previously parametric lens models of AS1063 in \cite{Bergamini2019} and \cite{Limousin2022}. Both models rely on a PIEMD parameterization for the main DM halo, and both find a similar PIEMD core radius of 90 kpc \citep[see also][who finds a similar PIEMD core radius]{Granata2022}. For these models, we only plot the main halo component, which is the main contributor to the lensing mass, especially beyond $r>10"$.

The central surface density for the Burkert profile, $\Sigma_o=r_b\rho_o$, is $\Sigma_o=2475 \Msun\, {\rm pc}^{-2}$. This is about an order of magnitude larger than the alleged universal value found in galaxies and dwarfs \citep{Donato2009}, and about a factor three times higher than the value found in groups \citep{Gopika2021}, thus confirming the non-universality of $\Sigma_o$  \citep{Zhou2020}, and that the value of $\Sigma_o$ grows with halo mass. 
Following \cite{Bondarenko2018}, we compute the average density $<\rho>$, and surface density $\Sigma(r)$, for the two analytical profiles. For the NFW and at the NFW scale radius, $r_s=300$ kpc, we find $<\rho>=0.00246\, \Msun {\rm pc}^{-3}$, and for the surface density (see their Eq. 3.2) we find $\Sigma_{\rm NFW}(r_s)=735\, \Msun {\rm pc}^{-2}$. Similarly, for the Burkert profile and within the smaller core radius of 150 kpc we find $<\rho>=0.0197\, \Msun {\rm pc}^{-3}$ and $\Sigma_{\rm B}(r_b)=946\, \Msun {\rm pc}^{-2}$. 
The values of $\Sigma(r)$ are in good agreement ($1\sigma$) with the scaling law (Eq. 3.6 in that paper) for a halo with virial mass of $2\times10^{15}\, \Msun$.

The fact that a similar cored profile in the distribution of DM is also found in the independent models of \cite{Bergamini2019}, \cite{Limousin2022}, and \cite{Granata2022} is reassuring. The PIEMD model has apparently a smaller core (90 kpc) than the Burkert model (150 kpc), but this is due to the different definition of a core in the PIEMD and Burkert profiles. Despite this difference in definition, as shown in Fig.~\ref{Fig_Profiles_V4} both the Burkert (150 kpc core) an PIEMD (90 kpc core) profiles are in excellent agreement with each other in the region constrained by lensing. Here we use the Burkert core of 150 kpc when we refer to the DM core radius. We also fit the GC (black line) profile with a Burkert profile and find a smaller core of 70 kpc. The Burkert model is shown in this case as a solid orange line in Fig.~\ref{Fig_Profiles_V4}. The fit is not perfect but is sufficient for our purposes of comparing core sizes obtained with the same ruler, the Burkert profile.  The size of the core radius for the GC distribution is significantly larger than the core in the GC distribution of nearby (but less massive) clusters, where given their low redshift GCs can be more easily detected. For instance, Coma ($r_b\approx 20$ kpc) or Virgo ($r_b\approx 5$ kpc) \citep{Peng2008,Peng2011} have much smaller cores for their GC system, suggesting that the core radius of the GC system may scale with the halo mass. 

To put the DM core in context, we follow \cite{Hayashi2025}. We computed the mean surface mass density within $r_{0.01}=0.01r_{v_{max}}$, where $r_{v_{max}}$ is the radius at which we find the maximum circular velocity. We compute the maximum circular velocity by the usual $V_c=\sqrt{GM(<r)/r}$ from the NFW profile shown in Fig.~\ref{Fig_Profiles_V4} (light blue curve) and find a maximum circular velocity ${v_{max}}=2115$ km s$^{-1}$ at $r_{v_{max}}=132"$ (in agreement with the expected $r_{v_{max}}=2.163r_s$ in NFW halos, with $r_s=300$ kpc for our NFW model). For comparison, using the Burkert profile \citep{Burkert1995}, shown as a red solid line in the same figure, we find ${v_{max}}=2083$ km s$^{-1}$ at $r_{v_{max}}=99"$. We adopt the NFW value as the most conservative since it gives a higher $r_{0.01}=1".32$. For the NFW profile we find $\Sigma(r<r_{0.01})=2529\, \Msun$ pc$^{-2}$ while for the Burkert profile we find  $\Sigma(r<r_{0.01})=143\, \Msun$ pc$^{-2}$. The value for the Burkert profile (that fits better the lens model) places this density as an extreme value, below all clusters with ${v_{max}}>1500$ km s$^{-1}$ listed in the recent compilation of  \cite{Hayashi2025}  \citep[see end of their Table 2, see also][]{Umetsu2016}. The value of  $\Sigma(r<r_{v_{max}})=143\, \Msun$ pc$^{-2}$ is closer to the theoretical model in \cite{Kaneda2024} of a halo that has experienced cusp-to-core transition (see their Figure 4). 

The fact that the DM core is approximately twice as large as the core observed in the distribution of GCs is intriguing if we consider that both GCs and DM are expected to behave as collisionless particles. Dynamical friction should be more effective on GCs than on the alleged subatomic DM particles and may play some role on the higher concentration observed in the GC distribution. However, the presence of a core in the center would make dynamical heating by GCs highly inefficient (see discussion about core stalling in Sec.~\ref{sec_CDM}). 


Unlike the cored distribution of GCs, the profile of satellite galaxies (or substructure) shows no core in N-body simulations \citep{Moore1999a,Nelson2024}. 
The radial distribution of satellite galaxies in Illustris-TNG in \cite{Nelson2024} shows no core around the center of the simulated or observed clusters. 
Based on real observations, data from the GAMA survey reveals small cores in the distribution of satellite galaxies of radius $\lesssim 20$ kpc at the center of clusters \citep{Riggs2022}. In the same work, the authors find that in the range $20\, {\rm kpc} < r < 500\, {\rm kpc}$ the number density of satellite galaxies falls as $\sim r^{-1}$. 
No core is observed either in the distribution of DM in the ultrahigh resolution NewCluster simulation \cite{Han2026,Jeon2025}. 

As mentioned in the introduction, lack of N-body simulations that can simultaneously simulate massive $10^{15}\, \Msun$ galaxy clusters and resolve small $10^5\, \Msun$ GCs hinders our ability to properly contextualize the difference in core radii between the GC and DM distributions. Before discussing the different scenarios and earlier work that can help us get some insight into the reason for the difference in core radii, we briefly discuss the dynamical state of AS1063. 

\section{Dynamical state of AS1063}

X-ray Chandra data reveal a very hot cluster ($kT>11.5$ keV). Based on the morphology of the X-ray emission and the high temperature, AS1063 was originally interpreted as a postmerger cluster \citep{Gomez2012}. The postmerger scenario is also suggested by studies of the intracluster light \citep{deOliveira2022},  and more recently by a joint lensing, X-ray and kinematics study in \cite{Beauchesne2024}. The postmerger scenario is supported as well by the presence of a radio halo \cite{Xie2020}, with a steep spectrum more consistent with turbulence re-acceleration (merger induced turbulence re-accelerates cosmic ray electrons) than a hadronic model (proton-proton collisions re-accelerates cosmic ray electrons) \citep{Rahaman2021}. A detailed analysis combining X-ray and radio data finds evidence for both a postmerger and a more relaxed scenario with some observational features resembling a cold core typical of relaxed clusters \citep{Rahaman2021}. 
This suggests that AS1063 has experienced a merger in the past (but not recently) and it is currently in the process of reaching equilibrium.\\

\begin{figure} 
  \includegraphics[width=9cm]{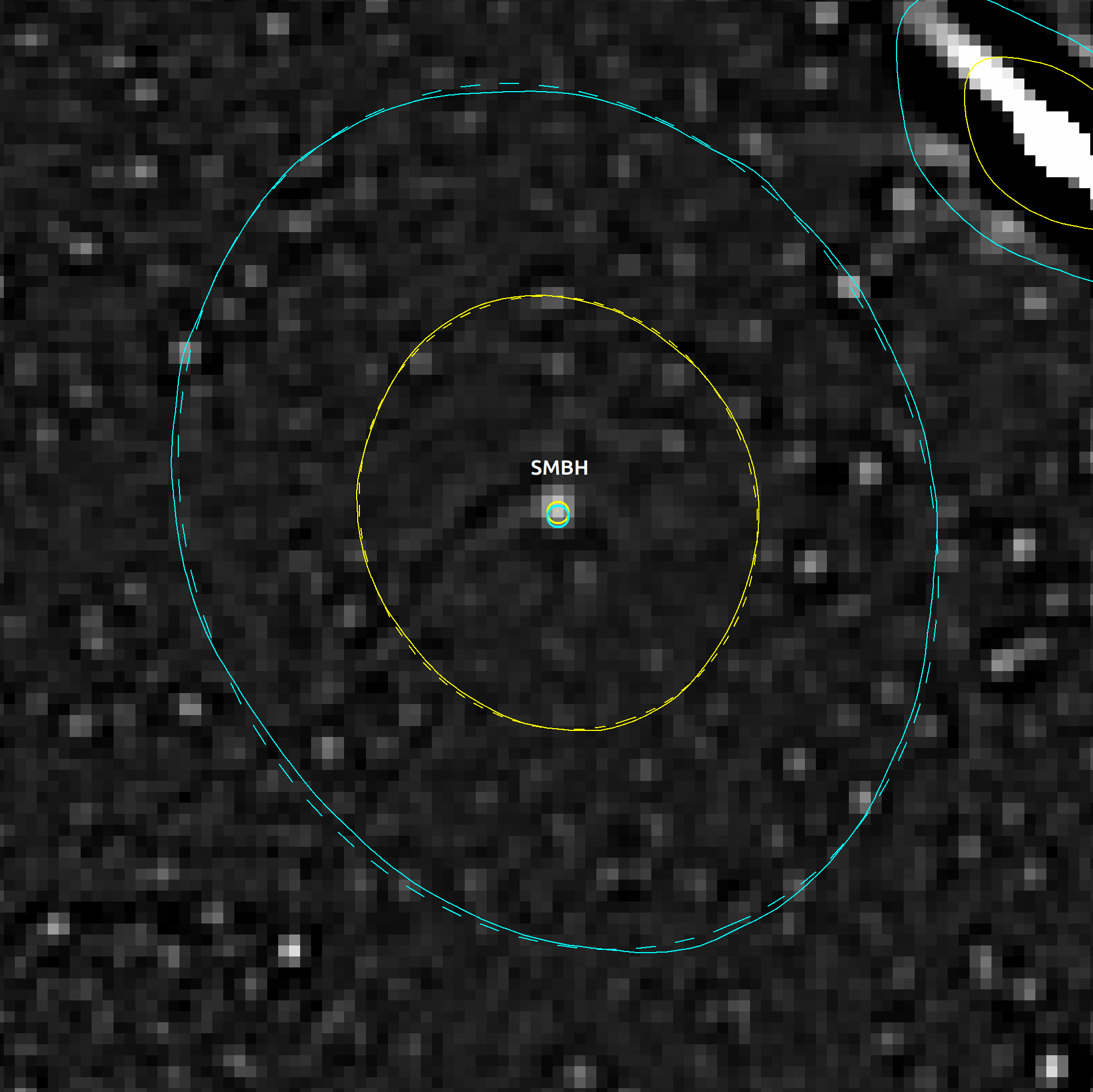} 
      \caption{Central $3"\times3"$ region of AS1063 (filtered). The image is obtained after two high-pass filters are applied to the F090W, F115W, F150W and F200W images \citep[see][for details]{Diego2026c}, and combining the four filtered images into a single one. Many GCs can be seen in the image with an almost uniform number density.  The yellow and cyan solid lines are the light contours in the raw F356W image at $50\%$ and $25\%$, respectively,  of the maximum intensity in the center. The yellow and cyan dashed curves are ellipses fitting the corresponding solid lines. The small yellow and cyan circles near the bright central source are the corresponding centers of the two ellipses. The bright source near the center is at the position of the peak emission in the raw images and it is likely the central SMBH.  
         }
         \label{Fig_CentralRegion}
\end{figure}

We can get further evidence from the JWST images. The isochrones around the BCG at $25\%$ and $50\%$ the peak flux at the center of the BCG are very symmetric and well described by ellipses (dashed lines in Fig.~\ref{Fig_CentralRegion}). The center of these ellipses are perfectly aligned with the peak emission at the BCG's center. 
To highlight small structures in the central region, we applied the same filtering process to the short-wavelength bands of JWST images as in \cite{Diego2026d}. The resulting filtered image is shown in Fig.~\ref{Fig_CentralRegion}. At the very center of the BCG, we find a compact source which coincides with the peak emission in JWST images. This source is allegedly the SMBH expected at the minimum of the potential well. The SMBH is found less than 30 milliarcseconds away from the center of the ellipse models fitting the light distribution. Given the lack of offset between the center of the light emission (or stellar component) and the SMBH, and the nearly perfect elliptical shape of the isochrones, we find no evidence of sloshing of the stellar component around the minimum of the potential,  which would have pointed to a recent merger episode. Hence, we conclude that the cluster central region is already (or close to being in) a relaxed state. \\
Finally, and relevant for the discussion in the next section, the filtered JWST images show no evidence of a resolved binary SMBH that could play a role on heating up the central region \citep[scouring by a binary SMBH, see][]{Postman2012}. However, this possibility can not be ruled out. While the detected SMBH may have emission around it from a compact cusp of stars (dress) or a small accretion disk, the companion SMBH (invisible in JWST images) may be a naked (no dress) and quiescent SMBH. 

%
%
%
%
%

\section{The case for CDM}
\label{sec_CDM}

The CDM scenario has been studied extensively with N-body simulations. In general these simulations predict a cusp at the center of massive halos \citep{NFW,Moore1999b,Springel2005,Diemand2005,Stadel2009,Navarro2010,Laporte2015,Haggar2021,Ishiyama2021}. This is in tension with the large cores observed in massive clusters such as AS1063. It is unclear how to produce large cores in galaxy clusters in the context of CDM. On smaller galaxies, such as dwarf galaxies, the shallower potential makes it easier for stellar feedback processes (supernova-driven outflows) to produce large cores  \citep{Mashchenko2008,deBlok2010,Chan2015,Read2019}, while for larger halos this mechanism is less efficient \citep{DiCintio2014,Tollet2016,Lazar2020,Dekel2021}.

As discussed in the previous section, AS1063 shows signs of a past merger, although it also shows signs of a near-equilibrium state. Mergers are a possible mechanism to produce cores, especially after two SMBH sink to the center of the halo forming a SMBH binary, inducing three body interactions, or SMBH scouring. This mechanisms can efficiently expel lighter material (stars and DM) from the central region \citep{Makino1996,Goerdt2010,Postman2012,Bonfini2016,Nasim2021,Khonji2024}. However, this effect is often restricted to small scales (few kpc on cluster scales) and does not explain why the distribution of light (stars) shows a much more compact configuration (core radius of few kpc, see Fig.~\ref{Fig_CentralRegion}) than the radial distribution of GCs or DM. 
Extrapolating the results from \cite{Rusli2013}, a core size of 150 kpc would imply a SMBH mass of at least $\approx 8.5\times10^{11}\, \Msun$ (using the most conservative reverse model in table 5 of that reference), or more than three times the expected upper limit for an accreting (and rotating) SMBH \citep{King2016}, although SMBH more massive than this limit can of course form via mergers. 

An alternative way of looking at mergers is by focusing on clusters that have experienced a recent violent merger. Perhaps one of the most extreme cases is the Bullet cluster \cite{Markevitch2004} given its mass, high relative velocity between groups, and small impact parameter during the interaction. Recent lens models have reconstructed the inner profile with great accuracy.  Both \cite{Cha2025,Rihtarsic2026} find that the inner 50 kpc profiles are cuspy, not flat, for both the main cluster and the bullet subgroup. 
Perhaps the Bullet cluster is observed too close to the crossing of the two groups and not enough time has passed for relaxation to modify the mass profiles. An interesting insight is given by \cite{Faltenbacher2006}, which traces the central density during a major merger as a function of time. As expected, there is an oscillatory pattern in the central density as the two groups cross each other, but the most distinguishing feature is a momentary increase by $\approx 30\%$ (lasting a few hundred million years) in the central density, followed by a return to a value of the central density close to the one before the interaction. The most intense change takes place during the first peri-center passage, with the effect diminishing in subsequent peri-center passages. As the merger evolves, the general trend for the central density is to increase, not decrease \citep[see also][]{Zemp2008}. The final merged halo in \cite{Faltenbacher2006} has a central cusp, in agreement with other N-body simulations that also find that cusps survive cluster mergers, as we explore next.

Simulations of dry mergers of two clusters find that mergers are relatively ineffective at changing the central densities of halos.  In these simulations, the remnant remembers the density profiles of the two clusters before merging, with the profile of the remnant being cuspy if one (or both) of the two original clusters is cuspy, while a cored profile after a merger is obtained only if both clusters had a core before the merger \citep{BoylanKolchin2004,Dehnen2005,Kazantzidis2006,McMillan2007,Zemp2008,Nipoti2009,Vass2009,Drakos2019a,Drakos2019b}. Hence, in the merger scenario discussed earlier, a cored profile can only be obtained (in collisionless DM models) if the two groups had large cores before merging.  
Moreover, adiabatic contraction may result in even more cuspy halos than in DM-only simulations, compensating, and in some cases reversing, baryonic effects that can reduce the central density \citep{Peirani2017}. 
Mergers can momentarily redistribute DM near the center of the halo with the DM density profiles becoming shallower than NFW, but the total mass density profiles (baryons plus dark matter) still resemble an NFW \citep{Laporte2015}. \\
In contrast, wet mergers, combined with dynamical friction,  may be more efficient at flattening the inner slope of the profile, although this has been studied in detail only on galaxy scale mergers \citep{ElZant2001,Hashim2024}. Semianalytic work shows that mergers involving a very compact satellite plus AGN feedback can be effective at transforming cusps into cores \citep{Dekel2021}. This view is challenged by N-body simulations of wet mergers where large amounts of cold gas can dynamically heat the DM cusp and transform it into a core, but keeping the profile of the total mass cuspy as the baryons replace the displaced DM at the center \citep{Ogiya2022}. 
On very massive galaxy clusters, it is unclear how this mechanism would work since cold gas can not survive the extreme (hot) environment of galaxy cluster cores. Runaway cooling flows at the very center are efficient at bringing large amounts of gas into the cluster center, that can later feed the central AGN providing feedback to heat up the core, although the AGN itself may heat up the infalling gas and stop the cooling flow \citep{Li2015,Guo2018}. Assuming AGN activity does not stop the inflow of cold gas, together with dynamical friction, AGN feedback is able to flatten the inner region, but this results in cores a few tens of kpc in size at most \citep{ElZant2004,DelPopolo2009,Governato2012,Martizzi2013}, significantly smaller than the 150 kpc core observed in AS1063. Hence, mergers and feedback seem insufficient to explain the large core observed in AS1063. Other dynamical effects must be at play in the case of CDM to heat up the central region of the cluster. \\

As mentioned in the discussion, one of the standing issues is the lack of N-body simulations that have the sufficient resolution to resolve the smallest substructures present in the cluster, in particular the GCs. Perhaps the abundant population of compact GCs (orders of magnitude more abundant than the subhalos in current N-body simulation) is efficient at dynamically heating the cusp and transforming it into a large core. Despite the lack of simulations on galaxy cluster scales, we can gain some intuition from simulations on smaller halos.  On galaxy scales, \cite{Boldrini2021} finds that accretion of satellites with highly eccentric orbits can heat up the central parts of halos, causing an outward migration of DM particles and transforming the central cusp into a core \citep[see also][for a similar study but with primordial black holes]{Boldrini2020}.
On even smaller scales, simulations of star clusters with different star masses (and remnants) do predict the formation of a core in star clusters, which agrees well with observations  \citep{Mackey2003,Baumgardt2017}. Particularly interesting for us are the simulations where a fraction of the objects are significantly more massive than  the rest. For instance,  black hole (BH) remnants from massive stars. \cite{Mackey2008} \citep[see also][]{Mackey2007} simulates star clusters with a population of stellar BHs with masses $\approx 20$ times larger than the typical mass of the rest of regular stars in the GC. In that case the regular stars play the role of the much lighter DM particle while the BHs play the role of the massive GCs in AS1063. In those simulations it is observed how an initial core radius of 1.9 pc (Runs 1 and 2) increases to $\approx 7$ pc due to dynamical heating from the BHs. It is tempting to think that in AS1063 a similar mechanism is taking place with the numerous GCs in the central region heating up the core of the much lighter DM particles. This possibility begs the question of why such phenomenon is not observed in CDM N-body simulations of massive galaxy clusters that contain thousands of subhalos that could also heat up the central cusp and transform it into a larger core. Maybe the answer lies again in the limited resolution of these simulations, since the number of GCs that needs to be considered is approximately 2 orders of magnitude larger than the number of subhalos included in typical N-body simulations. In addition, CGs are much more compact than the haloes in CDM simulations and hence are able to survive close encounters with the center of the galaxy cluster and remain in close orbits around the halo center, instead of evaporating into the main halo. Although this is an interesting possibility that should be addressed with future simulations, the results from  \cite{Mackey2008} can not be taken at face value for a galaxy cluster with a system of $\sim 10^5$ GCs, since mechanisms such as two-body relaxation, or mass segregation, do not take place (or are much less efficient) in galaxy clusters than in star clusters. For instance, mass segregation is not observed in the distribution of GCs in AS1063. This is shown in Fig.~\ref{Fig_Profiles_V4} as a dashed line, which represent the number density of bright (or more massive) GCs, re-scaled by the same factor 4. This profile corresponds to the brightest (i.e., more massive) 4061 GCs in our catalog (or approximately one seventh the full sample). This profile is well described by the same Burkert profile with core radius 70 kpc, but divided by a factor seven.  If mass segregation was taking place, we would expect this profile to have a noticeably smaller core radius. The decrease in surface number density between $2"$ and $5"$ is interesting and statistically significant, especially given the fact that these objects are the brightest and easier to detect, so they are in principle not affected by incompleteness. Interpreting this feature is beyond the scope of this paper but it is intriguing enough to motivate future work. \\

In the classic CDM context, the large, but different, cores in the GC and DM distributions may represent an example of a "chicken and egg" situation. The similarity in profiles between the GC and DM distributions suggest that they are interconnected. The question is what core was created first, or is there any positive feedback mechanism between the two cores, making them both grow over time. The larger DM core could be a consequence of the GC system heating up the core, or the observed cored distribution of GCs could be tracing the larger DM core. Since dynamical friction becomes inefficient for GCs in a core \citep[core stalling][]{Read2006,Goerdt2010,Kaur2022,Banik2022,Modak2023,vandenBosch2025,DiCintio2025}, one could imagine the distribution of GCs as a history record of past mergers, with GCs near the center being accreted earlier (and remaining in nearly constant orbits due to core stalling), and the last accreted GCs remaining closer to the boundaries of the DM core which grows over time, as the already accreted GCs may prevent the core from transforming into a cusp. This is reasonable in the commonly accepted picture where the less gravitationally bound DM halo is being tidally stripped first from the infalling satellite galaxies, the more tightly bound GCs are removed afterwards and stars (occupying the minimum in the potential well of the satellite galaxies), are accreted last \citep{Smith2016,Martin2026},  ending up closer to the center of the cluster (i.e, the BCG). In order to derive a firm conclusion on the ability of CDM models to reproduce the observed cores in the CG and DM distributions of AS1063, the picture above should be tested with high-resolution N-body simulations that can resolve the small, but very numerous, GCs.

\section{The case for SIDM}
\label{sec_SIDM}
The large core of AS1063 is naturally explained in SIDM, which predicts the existence of large cores in galaxy, and  galaxy cluster scale halos \citep{Moore1994,Spergel2000,Burkert2000,Firmani2000,Rocha2013,Tulin2018,Robertson2019,Vega-Ferrero2021,Fischer2024}. These models differ from the standard CDM in the cross section, $\sigma_{\rm SI}$, which although small, is much larger than in CDM (where formally $\sigma_{\rm CDM}$ can be as low as zero). In principle,  $\sigma_{\rm SI}$ can depend on the velocity, which helps explain the existence of cores in low-mass dwarf galaxies and high-velocity galaxy clusters, while in the intermediate halo mass range of galaxies, gravothermal core collapse can take place, resulting in profiles that resemble cuspy halos. A velocity dependent cross section is expected in DM models that involve a scalar mediator or a Yukawa potential \citep{Dooley2016,Nadler2020,Correa2021,Ragagnin2024}, or if DM is made of dark atoms \citep{Cline2014}. 

A SIDM interpretation for this cluster offers a natural explanation for the puzzling situation presented by the smaller core observed in the distribution of GCs. Since GCs are expected to behave as truly non-interactive particles, their cross section should be smaller than for SIDM and they should be a good tracer of the gravitational potential. The smaller core in the GC distribution may be explained by the core stalling mechanism and continuous mergers  discussed at the end of the previous section. However, similarly to the CDM case, detail N-body simulations (this time in the SIDM scenario) resolving the large population of GCs in galaxy clusters are lacking, so it is not clear if a core radius for the GC distribution about half the size of the core radius for the DM distribution is still consistent in the SIDM case. 
The SIDM scenario agrees well also with the apparent deficit of member galaxies (subhalos) near the center of the halo, a characteristic that is easier to interpret with the stronger ram pressure stripping caused by self-interactions with the cluster in the case of SIDM \citep{Nagai2005,Nadler2020,Sirks2022}.

CDM seems to struggle also with lensing observations by subhalos that apparently are more lensing efficient than expected from simulations \citep{Meneghetti2020}. This can be explained in the context of SIDM where the increase in lensing efficiency of subhalos is due to gravothermal core collapse which is more likely to take place in the subhalos if $\sigma_{\rm SI}$ has a dependency with the velocity of the DM particles \citep{Fischer2025,Natarajan2026}. \\

Following \cite{Rocha2013}, we estimate the Burkert core radius for a SIDM model with $\sigma_{\rm SI}=1$ cm$^2$ g$^{-1}$:
\begin{equation}
r_b=7.5\,{\rm kpc}\left(\frac{v_{max}}{100\, {\rm km s}^{-1}}\right)^{1.31} =400\,\, {\rm kpc}\,.
\end{equation}
This is about a factor 2.7 times larger than the value obtained for AS1063 so we conclude that $\sigma_{\rm SI}<1$ cm$^2$ g$^{-1}$

Interpolating the core radii found in \cite{Rocha2013} for $\sigma_{\rm SI}<0.1$ cm$^2$ g$^{-1}$ and $\sigma_{\rm SI}<1$ cm$^2$ g$^{-1}$, we find that $\sigma_{\rm SI}\approx 0.3$ cm$^2$ g$^{-1}$ should produce a Burkert core size $r_b\approx 150$ kpc for a halo with $v_{max}\approx 2000$ km s$^{-1}$. This is comfortably below the limit set by the Bullet cluster of $\sigma_{\rm SI}\lesssim 1$ cm$^2$ g$^{-1}$ \citep{Markevitch2004,Randall2008}. 

While SIDM offers a plausible solution for AS1063, this does not solve the question of why some clusters show large cores and some do not. Due to the large velocities, and relatively late formation history, galaxy cluster scale halos are not expected to reach the gravothermal collapse phase during a Hubble time. 
Detailed lensing models in other clusters find a variety of slopes in the center of clusters, with some clusters resembling NFW-like profiles while other show large cores. 
To add to the confusion, some merging clusters exhibit significantly different profiles in each of the subclusters. For instance, in A370 the south subcluster has a cental DM cusp while the northern subcluster has a large core \citep{Diego2025a}, with the inner region ($r<20$ kpc) dominated by the stellar component. Perhaps the velocity dependence of $\sigma_{\rm SIDM}$ can offer a solution to this conundrum if the velocity dispersion can be reduced by some mechanism near the center of some halos, thus increassing the probability of interaction and reducing the central density, which in turn may reduce the velocity even further, resulting in a runaway process that flattens the core. This process could be followed by gravothermal collapse in some cases, similar to the mechanisms invoked for galaxy-scale halos, resulting in the formation of a new cusp.

\section{The case for $\psi$DM} 
\label{sec_psiDM}
Another model that predicts cores in the center of halos is $\psi$DM \citep{Peebles2000,Goodman2000,Hu2000,Schive2014a,Marsh2016,Schive2025}.
The size of the core is linked to the de Broglie wavelength of the $\psi$DM particle, $\lambda_{\rm dB}$, determined by the mass of the DM particle and and the halo mass. For a $10^{15}\, \Msun$ halo and $m_{\psi}=10^{-22}$ eV, $\lambda_{\rm dB}=15$ pc \citep{Broadhurst2025}. N-body simulations show how the size of the core scales with the halo mass as $R_c\propto M_h^{-1/3}$, resulting in a core for AS1063 of less than 1 kpc for the same model above \citep{Schive2014b}. A much smaller DM particle mass, $m_{\rm \psi}\sim10^{-26}$ eV, would be needed in order to explain the core on AS1063, but this mass is clearly ruled out by observations \citep{Marsh2016,Benito2025}. Hence, $\psi$DM does not offer a valid solution for the core of AS1063.

\section{Conclusions}\label{sect_conclusions}
We discuss the large cores found in AS1063 for its rich system of GCs as well as for the lensing mass (mostly DM). 
Thanks to the exceptional depth of JWST observations, and the  large number of lensing constraints in AS1063 within the core radius (approximately half the lensing constraints fall within 150 kpc), the existence and scale of the core radii for both GCs and DM are well established. 
We measure the core sizes for the DM and GC distributions and find that the DM has a core approximately twice as large as the GC distribution (150 kpc, vs 70 kpc). Since both GCs and DM are expected to behave as collisionless particles in the standard CDM model, the much larger core size in the lensing mass is puzzling. We discuss mergers, feedback, and dynamical effects and find that such large core for the DM distribution is difficult to explain in the standard CDM scenario. The obserevd DM core is also much larger than the one predicted in $\psi$DM models with axion mass $m_{\psi}\approx 10^{-22}$ eV, and would require a much smaller axion mass, already ruled out by other observations. 
In contrast, the existence of such a large core in the DM can be explained in SIDM models with a velocity dependent cross section. For the scale (velocity) of a galaxy cluster, a cross section $\sigma_{\rm SI}\approx 0.3$ cm$^{2}$ g$^{-1}$ predicts a DM core with roughly the same size as the observed one. Despite the apparent success of the SIDM model,  we also argue that our results are inconclusive because of the lack of high resolution N-body CDM simulations of galaxy clusters with a large number ($\sim10^5$) of  $10^5$--$10^6\ \Msun$ GCs, which prevents us from ruling out CDM. We argue that such large number of GCs, accreted continuously during the formation of the cluster, and together with core stalling that freezes the GCs orbits as  they enter the DM core, may be able to reproduce both, a large core for the GC distribution and an even larger core for the DM, with collisionless DM particles.

\begin{acknowledgements}
The author thanks G. Yepes for useful comments. J.M.D. acknowledges the support of projects PID2022-138896NB-C51 (MCIU/AEI/MINECO/FEDER, UE) Ministerio de Ciencia, Investigaci\'on y Universidades and SA101P24.

This work is based on observations made with the NASA/ESA/CSA \textit{James Webb} Space Telescope. The data were obtained from the Mikulski Archive for Space Telescopes at the Space Telescope Science Institute, which is operated by the Association of Universities for Research in Astronomy, Inc., under NASA contract NAS 5-03127 for JWST. 
These observations are associated with JWST program \#3293. 

\end{acknowledgements}

\bibliographystyle{aa} 
\bibliography{MyBiblio} 


\end{document}